\documentclass[12pt]{iopart}

\usepackage{setstack}
\usepackage{iopams}
\usepackage{amsthm}
\usepackage{graphicx}
\usepackage{amscd}

\theoremstyle{definition}
\newtheorem{Def}{Definition}[section]
\newtheorem{Thm}{Theorem}[section]

\newcommand{\V}{\mathbf{V}}
\newcommand{\pr}{\mathbf{pr}}
\newcommand{\p}[1]{\partial_{#1}}

\begin{document}%

\title{Point transformations in invariant difference schemes}

\author{Francis Valiquette}

\address{Centre de Recherches Math\'ematiques, Universit\'e de
      Montr\'eal, C.P. 6128, succ. Centre-ville, Montr\'eal, QC, H3C
      3J7, Canada}

\ead{valiquet@crm.umontreal.ca}

\begin{abstract}
In this paper, we show that when two systems of 
differential equations admitting a symmetry group
are related by a point transformation
it is always possible to generate invariant schemes, 
one for each system, that are also related by the same transformation.
This result is used to easily obtain new invariant schemes
of some differential equations.   
\end{abstract}

\submitto{\JPA}

\section{Introduction}%

In modern numerical analysis, the development of 
geometric integration has become an increasingly
active field of research.  Geometric integration
is related to the development of new numerical schemes 
that incorporate additional structures of the dif\mbox{}ferential
equation which is being discretized in order to
reproduce the qualitative features of the continuous solution,
\cite{HLW}.  

For dif\mbox{}ferential equations possessing symmetries,
a natural thing to ask for, when discretizing such equations,
is to preserve as much symmetry as possible.  
There are mainly two approaches of dealing with such a problem.  
The first one consists of defining invariant schemes
on fixed lattices, \cite{FNNV,LVW,LTW-2004}, and consider 
only transformations that do not act on the lattices.  
In order to obtain interesting
symmetries, transformations acting simultaneously 
on more than one point of
the mesh must be considered.  A second point of view 
consists of defining invariant numerical schemes 
over evolutive lattices.
Such an approach is used when considering groups of 
transformations acting both
on the dependent and independent variables.  
In the latter case, two dif\mbox{}ferent methods 
of addressing the issue 
are found in the literature. 
The first one is based on the application of the moving frame method,
\cite{KO,O-2001}, while the other uses a method
based on the infinitesimal generators of symmetries,
\cite{DK,DKW-2000,DKW-2004,RW}.  
In this paper, we shall be concerned only with invariant schemes 
generated using the last approach.

The applications of the Lie groups to discrete equations
is quite recent compared to its continuous conterpart, which
goes back to the work of Sophus Lie, \cite{L-1,L-2}.  The first 
articles in this new field appeared at the beginning of the
1990-ties, \cite{D-1991,LW-1991}.  The goal is to develop the 
applications of the Lie groups to discrete equations
into tools that will be as powerful as for dif\mbox{}ferential equations.
The interested reader
can find a review of important results and an extensive bibliography
in \cite{LW-2005}.  

The aim of this work 
is to systematize a result obtained in a paper of Dorodnitsyn and
Kozlov, \cite{DK-1997}.  In the article it shown that the
invariant schemes for the one dimensional
Burgers' equation in the potential form, $w_t+\frac{1}{2}
w_x^2=w_{xx}$, are related to the invariant schemes for the one dimensional
linear heat equation, $u_t=u_{xx}$, by the point transformation
$w=-2\ln(u)$, the same transformation relating the continuous
equations.  We shall see that this is just an
example of a more general result.  It is well known that when two
realizations of
Lie symmetry algebras of dif\mbox{}ferential equations 
by vector fields can be transformed into each other by
a point transformation, then the same transformation will 
map the two dif\mbox{}ferential equations one to the other.  In fact, 
this result is also true for the invariants schemes approximating
the dif\mbox{}ferential equations.  In other words, when to dif\mbox{}ferential 
equations are related by a point transformation, it is possible
to define invariants schemes for the two equations that 
are also related by the same
transformation.  This result has some interesting applications.
For example, it can happen that the discretisation of a dif\mbox{}ferential
equation is easier to find if a change of coordinates is performed.
If that is so, one can then generate an 
invariant scheme in the new variables and find the desired result
by performing the inverse transformation.  A second direct application
consists of obtaining solutions of an invariant scheme from known solutions 
of a related invariant scheme.

The article can be outlined in the following way.  First of all, 
we recall the algorithm for generating
invariant schemes for a system of dif\mbox{}ferential equations. 
It is possible to generate symmetry-preserving schemes
of systems of ordinary dif\mbox{}ferential equations (ODEs) as well as
systems of partial dif\mbox{}ferential equations (PDEs) with the algorithm.  
Secondly, we show that when two systems of dif\mbox{}ferential equations are 
related by a point transformation, the transformation will
also relate the invariant schemes for the two systems.
Finally, we apply the result to dif\mbox{}ferent types of point transformations.

\section{Invariant Dif\mbox{}ference Schemes}%

Let $x=(x^1,\ldots,x^p)\in X=\mathbb{R}^p$, $p\in \mathbb{N}$,
be the independent variables 
and $u:X\to U\subset\mathbb{R}^q$, $x\mapsto u(x)=(u^1(x),\ldots,u^q(x))$, 
the dependant variables of the
system of dif\mbox{}ferential equations
\begin{equation}
\Delta(x,u^{(n)})=0\label{DE}.
\end{equation}
In the equation \eref{DE}, $u^{(n)}$ denotes all the 
derivatives of $u(x)$ up to order $n$ with respect to 
all the independent variables.  
Since we want to generate invariant numerical
schemes, we suppose that \eref{DE}
is invariant under a group of Lie point symmetries $G$, 
of order $N$, generated by the vector fields
\begin{equation}
\V_k=\sum_{i=1}^p\xi^{k,i}(x,u)\p{{x^i}}
+\sum_{\alpha=1}^{q}\phi^{k,\alpha}(x,u)\p{u^\alpha},\qquad 
k=1,\ldots, N.\label{symmetry generators}
\end{equation}
The set $\{\V_k\}$ forms a basis of a Lie algebra $L$.

The discretisation of \eref{DE} consists of sampling in 
the space of independent variables $X$
some points $x$, each labelled by a set of discrete indices,
\begin{equation}
x_m=(x^1_m,\ldots,x^p_m),\qquad m\in \mathbb{Z}^p. 
\end{equation}  
The discretisation of the independant variables induces
a natural discretisation of the dependent ones: 
\begin{equation}
u_m=(u^1_m,\ldots,u^q_m).
\end{equation}
The construction of an invariant numerical scheme consists of
determining a consistent way of sampling the points $x_m$
and defining the evolution of the solution $u_m$
so as to preserve the symmetries of the continuous 
system of equations.  Since the symmetry transformations act on 
independent and dependent variables, we must let the symmetry group
$G$ act both on the finite dif\mbox{}ference equations
approximating the system of dif\mbox{}ferential equations and the
lattice on which the approximation is made.  

To adequately model the discrete problem 
we consider a system of 
$q+p$ finite dif\mbox{}ference equations
\begin{equation}
\fl E_k(\{(x_{m+j},u_{m+j})\}_{j\in J})=0,\qquad
1\leq k \leq q+p,\qquad {\rm where}\;\{0\}\subset J
\subset \mathbb{Z}^p,
\label{sfde}
\end{equation}  
relating the quantities $(x,u)$ at a finite
number of points. The set $J$ in \eref{sfde}
is of finite order and serves to identify the neighbouring
points of $(x_m,u_m)$. 
By choice, we suppose that the first $q$
equations approximate the system of dif\mbox{}ferential equations while the
last $p$ equations specify the mesh.  The $p$ equations for the
lattice do not completely determine the lattice in general.  In fact, they
impose restrictions on it.  
The number of points related with each other in \eref{sfde} 
depends on the order of the original dif\mbox{}ferential
equation and the precision we look for. 
In the continuous limit, we impose that the first $q$ 
equations of \eref{sfde} go to the system of
dif\mbox{}ferential equations while the $p$ other go
to the identity $0=0$.  Finally, to have an
invariant scheme, we request that the system \eref{sfde} 
be invariant under the group of transformations generated
by \eref{symmetry generators}.  

The procedure for generating invariant finite 
dif\mbox{}ference equations from a known symmetry group is
analogous to the continuous case.  As for the
continuous case, we define a prolongation of
the group action, but in a dif\mbox{}ferent fashion,
\cite{LW-2005,DK,DKW-2000,DKW-2004,RW,DK-1997,LTW}.
The prolongation of the group action for the discrete
problem is realized by requiring that the group acts
on all points figuring in \eref{sfde}
\begin{Def}
Let $G$ be a group of point transformations acting on
the space $X \times U$.  The discrete prolongation
of the group action is defined as
\begin{equation}
\pr\; g\cdot\{(x_{m+j},u_{m+j})\}_{j\in J}
=\{g\cdot(x_{m+j},u_{m+j})\}_{j \in J},\qquad\forall g\in G.
\end{equation}
\end{Def}
In fact, the most general definition of the discrete prolongation
of the group action would require that the transformation acts simultaneously
on all the points of the discrete space.  Since the
system of dif\mbox{}ference equations \eref{sfde} only involves a finite
number of discrete points, we can restrict the prolongation
of the action to those points. 
With this definition of the discrete prolongation of the group
action, it is straightforward to derive an expression
for the discrete prolongation of an infinitesimal generator of 
tranformations.
\begin{Def}
Let $\V_k$ be vector fields, as in  \eref{symmetry generators}
defining a basis of $L$.
The discrete prolongation of $\V_k$ is defined as
\begin{equation}
\pr\;\V_k:=\sum_{j\in J}[\sum_{i=1}^n\xi_{m+j}^{i,k}\p{{x_{m+j}^{i}}}
+\sum_{\alpha=1}^q\phi_{m+j}^{\alpha,k}\p{{u^\alpha_{m+j}}}],
\label{discrete prolongation}
\end{equation}
where $\xi_{m+j}^{i}=\xi^{i}(x_{m+j},u_{m+j}) \;{\rm and }\; 
\phi_{m+j}^\alpha=\phi^\alpha(x_{m+j},u_{m+j})$.
\end{Def}

The method for generating a set of fundamental invariants in the
discrete case is identical to the continuous one, 
\cite{O-1993,O-1995}.  The only dif\mbox{}ference is that instead
of using the continuous prolongation of the group action
we use the discrete prolongation.  In the discrete
situation, an invariant involving the points
$\{(x_{m+j},u_{m+j})\}_{j\in J}$is a quantity that satisfies
\begin{equation}
I(\pr\; g\{(x_{m+j},u_{m+j})\}_{j\in J})
=I(\{(x_{m+j},u_{m+j})\}_{j\in J}), \qquad \forall g\in G.
\label{invariance condition}
\end{equation}

Hence, given a basis of the Lie symmetry algebra $L$, 
\eref{symmetry generators}, we look for the quantities $I$
satisfying 
\begin{equation}
\pr\; \V_k [I(\{x_{m+j},u_{m+j}\}_{j\in J})]=0,\qquad k=1,\ldots,N.
\end{equation}
Using the method of characteristics, 
we obtain a set of elementary invariants
$I_1,\ldots,I_\mu$.  Their number is given by the formula
\begin{equation}
\mu={\rm dim}\; M - {\rm rank}\; Z,
\label{number of invariants}
\end{equation}
where $M$ is the manifold that $G$ acts on, i.e. 
$M\sim\{\{x_{m+j},u_{m+j}\}_{j\in J}\}$.  So
${\rm dim}\;M= p+q\times \#J$,  where $\#J$ denotes 
the order of the set $J$.  
$Z$ is the $N\times (p+q)\times \#J$ matrix 
\begin{equation}
Z=
\left(\begin{array}{c}
\{\xi^{1,1}_{m+j},\xi^{1,2}_{m+j},\ldots,\phi^{1,q-1}_{m+j},
\phi^{1,q}_{m+j}\}_{j\in J}\\
\vdots\\
\{\xi^{N,1}_{m+j},\xi^{N,2}_{m+j},\ldots,\phi^{N,q-1}_{m+j}
\phi^{N,q}_{m+j}\}_{j\in J}
\end{array} \right)
\end{equation}
formed using the coef\mbox{}ficients 
of the prolonged symmetry generators \eref{discrete prolongation}.
Since the quantities $I_1,\ldots,I_\mu$ form a basis
of elementary invariants, any
invariant dif\mbox{}ference 
equation must be written as
\begin{equation}
E(I_1,\ldots,I_\mu)=0.\label{strong invariant}
\end{equation}
Equations \eref{strong invariant} obtained in this 
manner are said to
be strongly invariant and satisfy $\pr\; \V_k [E]=0$ 
identically, \cite{LW-2005}.

Further invariant equations can be obtained if the rank of the matrix 
$Z$ is not maximal on some manifolds described by equations of the
form $E(\{x_{m+j},u_{m+j}\}_{j\in J})=0$ and
satisfy
\begin{equation}
\pr\; \V_k[E]\bigg\vert_{E=0}=0,\qquad k=1,\ldots,N.
\label{weakly invariance} 
\end{equation}
Such equations are said to be weakly 
invariant, \cite{LW-2005}.  
In practice we usually start by computing the 
invariant manifolds since
it can facilitate the computation of the 
set of fundamental invariants
afterwards.  In order for the system of 
finite dif\mbox{}ference equations 
\eref{sfde} to be invariant under the group of symmetries $G$,
it must  be formed out of weakly or strongly invariant 
dif\mbox{}ference equations and so each equation
will satisfy \eref{weakly invariance} on the space of
solutions.  Many different invariant schemes can be formed
using the obtained invariants.  The only requirement is
that the continuous limit of the invariant scheme
gives back the system of dif\mbox{}ferential equations we
are discretizing.

\section{Point transformations for invariant schemes}
\label{section point transformation}

In this section, we enunciate the main result of
this article.

\begin{Thm}
Let $\Delta(x,u^{(n)})=0$ and 
$\widetilde{\Delta}(\widetilde{x},\widetilde{u}^{(n)})=0$
be two systems of dif\mbox{}ferential equations related
by an invertible point transformation 
\begin{equation}
\psi:X\times U\to \widetilde{X}\times\widetilde{U},\qquad
(x,u)\mapsto (\widetilde{x},\widetilde{u})=(\psi|_x(x,u),\psi|_u(x,u)),
\label{point transformation}
\end{equation}
that also relates their respective symmetry groups $G$ and 
$\widetilde{G}$.
Then under the point transformation \eref{point transformation}
invariants schemes of $\Delta(x,u^{(n)})=0$ are mapped
to invariant schemes of 
$\widetilde{\Delta}(\widetilde{x},\widetilde{u}^{(n)})=0$.
\end{Thm}

\begin{proof}

Let 
\begin{equation}
\{I_1,\ldots,I_\mu\},
\label{original invariants}
\end{equation}
be a basis of discrete invariants under $G$ and
\begin{equation}
E_k(\{(x_{m+j},u_{m+j})\}_{j\in J})=0,\qquad k=1,\ldots,p+q,
\label{original FDE} 
\end{equation}
a invariant scheme of $\Delta(x,u^{(n)})=0$.
We want to show that under the point transformation 
\eref{point transformation},
the invariant scheme \eref{original FDE} is
mapped to an invariant scheme of the
system of differential equations 
\begin{displaymath}
\widetilde{\Delta}(\widetilde{x},\widetilde{u}^{(n)})
=\Delta(\psi|_x(x,u),(\psi|_u(x,u))^{(n)})=0.
\end{displaymath}

First of all, under the point transformation
\eref{point transformation} the set of points
$\{(x_{m+j},u_{m+j})\}_{j\in J}$ is mapped to
\begin{equation}
\{(\widetilde{x}_{m+j},\widetilde{u}_{m+j})\}_{j\in J}=
\{\psi(x_{m+j},u_{m+j})\}_{j\in J}.
\label{new discrete points}
\end{equation}
Hence, a set of elementary invariants of $\widetilde{G}$ 
relating the discrete points \eref{new discrete points} is given by
\begin{equation}
\widetilde{I}_k(\{\widetilde{x}_{m+j},\widetilde{u}_{m+j}\}_{j\in J})
=I_k(\{\psi^{-1}(\widetilde{x}_{m+j},\widetilde{u}_{m+j})\}_{j\in J}),
\qquad k=1,\ldots\mu,
\label{new invariants}
\end{equation}
where the $I_k$ are given in \eref{original invariants}.  Indeed,
using the fact that the relation between the symmetry groups $G$
and $\widetilde{G}$ is givent by 
\begin{equation}
\widetilde{g}=\psi \circ g \circ \psi^{-1}, \qquad\forall g\in G.
\end{equation} 
we see that the invariance condition \eref{invariance condition}
is satisfied by the quantities $\widetilde{I}_k$,
\begin{eqnarray}
\fl \widetilde{I}_k(\pr\;
\widetilde{g}\{(\widetilde{x}_{m+j},\widetilde{u}_{m+j})\}_{j\in J})
&=I_k(\{\psi^{-1}\circ \psi \circ g \circ \psi^{-1} \circ \psi 
(x_{m+j},u_{m+j})\}_{j\in J})\label{new invariance}\\
&=I_k(\{x_{m+j},u_{m+j}\}_{j\in J})
=\widetilde{I}_k(\{\widetilde{x}_{m+j},\widetilde{u}_{m+j}\}_{j\in J}),
\qquad \forall \;\widetilde{g}\in\widetilde{G}.
\nonumber
\end{eqnarray}
So all strongly invariant equations of \eref{original FDE}
are mapped by \eref{point transformation} to new strongly invariant
equations in terms of the discrete variables \eref{new discrete points}.

The same af\mbox{}firmation is true for weakly invariant equations.
Given a weakly invariant equation $E(\{(x_{m_j},u_{m+j})\}_{j\in J})=0$,
the equation
\begin{equation}
\widetilde{E}(\{(\widetilde{x}_{m_j},\widetilde{u}_{m+j})\}_{j\in J})=
E(\{\psi^{-1}(\widetilde{x}_{m_j},\widetilde{u}_{m+j})\}_{j\in J})=0
\end{equation}
is weakly invariant under $\widetilde{G}$.  Indeed,
\begin{eqnarray}
\fl\widetilde{E}(\pr\;\widetilde{g}
\{(\widetilde{x}_{m_j},\widetilde{u}_{m+j})\}_{j\in J})
\bigg|_{\widetilde{E}(\{(\widetilde{x}_{m_j},\widetilde{u}_{m+j})\}_{j\in
J})=0}\nonumber\\
\lo=E(\{\psi^{-1}\circ\psi\circ g 
\circ \psi^{-1}\circ\psi(x_{m_j},u_{m+j})\}_{j\in J})
\bigg|_{E(\{(x_{m_j},u_{m+j})\}_{j\in J})=0}=0.
\label{new weakly invariance}
\end{eqnarray}
So from \eref{new invariance} and \eref{new weakly invariance} we
conclude that the new system of finite dif\mbox{}ference equations
\begin{equation}
\fl\widetilde{E}_k(\{(\widetilde{x}_{m_j},\widetilde{u}_{m+j})\}_{j\in J})=
E_k(\{\psi^{-1}(\widetilde{x}_{m_j},\widetilde{u}_{m+j})\}_{j\in J})=0,
\qquad k=1,\ldots,p+q,
\label{new FDE}
\end{equation} 
is invariant under the group of symmetry $\widetilde{G}$.

The convergence of \eref{new FDE} to 
$\widetilde{\Delta}(\widetilde{x},\widetilde{u})=0$, in the
continuous limit, is verified since the diagram
\begin{equation}
\begin{CD}
\Delta(x,u^{(n)})=0 @>{\psi}>> 
\widetilde{\Delta}(\widetilde{x},\widetilde{u}^{(n)})=0\\
@A{\text{Continuous}}A{\text{limit}}A   
@A{\text{Continuous}}A{\text{limit}}A\\
E(\{x_{m+j},u_{m+j}\}_{j\in J})=0  @<<{\psi^{-1}}< 
\widetilde{E}(\{\widetilde{x}_{m+j},\widetilde{u}_{m+j})\}_{j\in J})=0 
\end{CD}
\end{equation}
commutes.

\end{proof}

Also, let us mention that 
any exact solution $u=f(x)$ of \eref{sfde}, which
means that $E(\{(x_{m+j},f(x_{m+j})\}_{j\in J})=0$ is satisfied
identically, is mapped by \eref{point transformation}
to an exact solution of \eref{new FDE} 
$\widetilde{u}=\psi|_u\circ f(\psi^{-1}|_x(\widetilde{x}))$ since
\begin{equation}
\widetilde{E}(\{(\widetilde{x}_{m+j},\widetilde{u}_{m+j})\}_{j\in J})=
E(\{\psi\circ\psi^{-1}(x_{m+j},f(x_{m+j}))\}_{j\in J})=0.
\end{equation}

\section{Applications}

In our applications, we restrict ourself to scalar dif\mbox{}ferential 
equations involving at most two independant variables.

To simplify the writing we introduce the following notation
\begin{eqnarray}
\fl(x^1_{m,n},x^2_{m,n},u_{m,n})\equiv(x^1,x^2,u),& &
(x^1_{m,n\pm 1},x^2_{m,n\pm 1},u_{m,n\pm 1})\equiv(x^1_\pm
,x^2_\pm,u_\pm),\\
\fl(x^1_{m+1,n},x^2_{m+1,n},u_{m+1,n})\equiv(\hat{x}^1,\hat{x}^2,\hat{u}),
& \qquad&
(x^1_{m-1,n},x^2_{m-1,n},u_{m-1,n})\equiv(\check{x}^1,\check{x}^2,\check{u}),
\end{eqnarray}
and introduce the steps
\begin{equation}
\Delta x^i\pm=\pm(x^i_\pm-x),\quad
\hat{\Delta x^i}=\hat{x}^i-x^i,\quad
\check{\Delta x^i}=x^i-\check{x}^i,\quad i=1,2.
\end{equation}

There are a lot of interesting point transformations that can
be considered.  We have chosen to look at three dif\mbox{}ferent
transformations.  The first application is concerned with
the hodograph transformation.  The transformation is used
to generate symmetry-preserving schemes of new equations and 
obtain exact solutions from known ones.  In the
second example we consider the wave equation in one
spatial dimension with a source term.  By reformulating
the problem in the characteristic variables, we shall see
that it is easy to derive an invariant scheme.  Then, by taking
the inverse transformation, an invariant
scheme in the original system of coordinates is obtained.  Finally, 
we investigate a particular example involving a change of variables
from cartesian to polar coordinates.

\subsection{The hodograph transformation}\label{hodograph}

Let us start by recalling the definition of a hodograph transformation
for a scalar dif\mbox{}ferential equation.

\begin{Def}
Let $x=(x^1,\ldots,x^p)\in \mathbb{R}^p$, and 
$u(x):\mathbb{R}^p\to \mathbb{R}$.  The transformation
\begin{equation}
\eqalign{
H:\mathbb{R}^{p+1}\to \mathbb{R}^{p+1}\nonumber\\
(x^1,\ldots,x^p,u)\mapsto (y=u,x^2,\ldots,x^p,v=x^1)
\label{hodograph transformation}}
\end{equation}
is called a pure hodograph transformation.

The quantity $v$ now plays the role of the independant variable.
Under such a transformation, the derivatives transform as 
\begin{eqnarray*}
u_{x^i}=-\frac{v_{x^i}}{v_y},\qquad i=2,\ldots,n,\\
u_{x^1}=\frac{1}{v_y}, \qquad
u_{x^1x^1}=-\frac{v_{yy}}{v_y^3},\qquad
u_{x^1x^1x^1}=-\frac{v_{yyy}}{v_y^4}+3\frac{v_{yy}^2}{v_y^5},
\end{eqnarray*}
and so on, \cite{A}.  
\end{Def}

This transformation
is found in the study of nonlinear dif\mbox{}ferential equations.  
Such a transformation is usually used to linearize 
dif\mbox{}ferential equations.  In our examples, we will
be interested in going the opposite direction.  Given a linear dif\mbox{}ferential
equation and its invariant scheme we use what we have seen
in Section \ref{section point transformation} to obtain  
an invariant scheme for the nonlinear dif\mbox{}ferential equation related
to the linear one by \eref{hodograph transformation}.   

\subsubsection{First Order Inhomogeneous Linear Equation}%

Let us consider the first order linear inhomogeneous ODE
\begin{equation}
u_x-A^\prime(x)u-B^\prime(x)e^{A(x)}=0.\label{linear ODE}
\end{equation}
This dif\mbox{}ferential equation admits the two 
dimensional symmetry algebra
\begin{equation}
\V_1=e^{A(x)}\p{u},\qquad \V_2=(u-B(x)e^{A(x)})\p{u}
\label{EDO linear algebra}
\end{equation}
and its general solution is
\begin{equation}
u(x)=(B(x)+c)e^{A(x)}\label{sol EDO lin},
\end{equation}
where $c$ is an integration constant, \cite{RW}.

To discretize \eref{linear ODE} we need one 
discrete variable, $m \in \mathbb{Z}$.  The schemes that
we will generate will involve the minimum number of points 
necessary to approximate the first order derivative.   This
means that it will only involve two points.

In order to generate an invariant scheme of \eref{linear ODE} 
we first start by finding a set of fundamental invariants
on the two-point scheme $\{(x,u),(x_+,u_+)\}$.  Hence we look
for the quantities that satisfy
\begin{equation}\eqalign{
(e^{A(x)}\p{u}+e^{A(x_+)}\p{u_+})I(x,x_+,u,u_+)=0,\\
((u-B(x)e^{A(x)})\p{u}+(u_+-B(x_+)e^{A(x_+)})\p{u_+})I(x,x_+,u,u_+)=0.}
\label{invariance condition linear ODE}
\end{equation} 
The solutions of \eref{invariance condition linear ODE} are
\begin{equation}
I_1=x,\qquad I_2=x_+.\label{invariants lin ODE}
\end{equation}
With these two invariants it is not possible
to obtain a symmetry-preserving scheme of \eref{linear ODE}.
However, the symmetry algebra \eref{EDO linear algebra}
admits the invariant manifold
\begin{equation}
u_+e^{-A(x_+)}-ue^{-A(x)}-B(x_+)+B(x)=0.
\label{weak inv lin}
\end{equation}
Hence an invariant scheme of \eref{linear ODE}
is given by the system of two equations
\begin{equation}
\eqalign{
u_+e^{-A(x_+)}-ue^{-A(x)}-B(x_+)+B(x)=0,\\
I_2-I_1=\epsilon,}
\label{invariant scheme linear ODE}
\end{equation}
where $\epsilon$ is a parameter that goes to zero in
the continuous limit.
This system of invariant dif\mbox{}ference equations forms
an invariant scheme of \eref{linear ODE}, since
in the continuous limit \eref{invariant scheme linear ODE}
goes to \eref{linear ODE}.  We recall that there are no recipes to
obtain \eref{invariant scheme linear ODE}.  The only requirements
are that the system \eref{invariant scheme linear ODE}
must be formed out of the
elementary invariants \eref{invariants lin ODE}
and the weakly invariant equation \eref{weak inv lin} and
that in the continuous limit we recover the ODE \eref{linear ODE}.

By substituing the continuous solution \eref{sol EDO lin}
into \eref{invariant scheme linear ODE} 
it is immediate to verify that it is
also an exact solution of the discrete problem \cite{RW}.  

\begin{figure}
\begin{center}
\includegraphics[height=8cm,width=10cm]{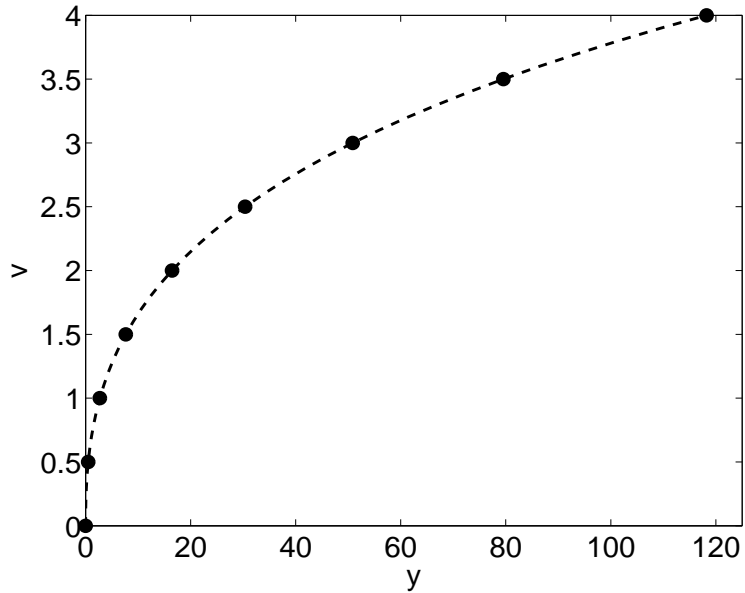}
\end{center}
\caption{Solution of
$v_y(1+4v^{3/2}\exp(\frac{1}{2}\sqrt{v}))=2\sqrt{v}$,\\
\begin{tabular}{cl}
\dashed & exact solution, \\
\fullcircle & discrete solution.
\end{tabular}
}
\label{numeric solution}
\end{figure}

Now, if we apply the hodograph transformation
$H:\mathbb{R}^2\to\mathbb{R}^2$, $(x,u)\mapsto
(y=u,v=x)$, equation \eref{linear ODE} transforms to the 
nonlinear ODE
\begin{equation}
v_y(A^\prime(v)y+B^\prime(v)e^{A(v)})-1=0\label{EDO linear-h}
\end{equation}
The symmetry generators of \eref{EDO linear-h} are
\begin{equation}
\widetilde{\V}_1=e^{A(v)}\p{y},\qquad
\qquad \widetilde{\V}_2=(y-B(v)e^{A(v)})\p{y}.\label{Lie alg lin h}
\end{equation}
If we apply the hodograph transformation to
\eref{invariant scheme linear ODE} we obtain
\numparts
\begin{eqnarray}
\widetilde{E}_1=y_+e^{-A(v_+)}-ye^{-A(v)}-B(v_+)+B(v)=0,
\label{lattice nonlinear ODE}\\
\widetilde{E}_2=v_+-v=\epsilon,\label{solution nonlinear ODE}
\end{eqnarray}
\endnumparts
and this system of equations satisfies the infinitesimal invariance
condition
\begin{equation}
\pr\;\widetilde{\V}_k[\widetilde{E}_i]\bigg|_{\widetilde{E}_1=0,
\widetilde{E}_2=0}=0,\qquad i=1,2,\quad k=1,2.
\end{equation}
When taking the continuous limit of \eref{lattice nonlinear ODE} and
\eref{solution nonlinear ODE} 
we recover from \eref{lattice nonlinear ODE} the nonlinear
ODE \eref{EDO linear-h} while \eref{solution nonlinear ODE} goes to
the identity $0=0$.  The invariant scheme 
obtained is quite dif\mbox{}ferent
from standard discretisation.  Indeed, in standard numerical
methods, the lattice can be chosen to have a variable step
size, but the choice does not incorporate any information on
the problem being discretized.  Usually, all the information is
incorporated in the finite dif\mbox{}ference equation approximating
the ODE.  In the invariant scheme, \eref{lattice nonlinear ODE} and 
\eref{solution nonlinear ODE}, we have the opposite situation.
Solving \eref{lattice nonlinear ODE} and \eref{solution nonlinear ODE} we find
that
\begin{equation}
v=m\;\epsilon+v_0\qquad\text{and}\qquad y=(B(v)+c)e^{A(v)},
\label{solution v}
\end{equation}
where $\epsilon$, $v_0$ and $c$ are constants.  The second
equation of \eref{solution v} defines the evolution
of the mesh in such a way that the dif\mbox{}ference
in $v$ between two iterations is constant. 
Hence, all the information of the continuous problem is
now incorporated entirely in the definition of the mesh.
In Fig. \ref{numeric solution}, we have plotted a particular
solution to illustrate the situation.  

\subsubsection{One Dimensional Linear Heat Equation}%

The one dimensional linear heat equation 
\begin{equation}
u_t=u_{xx} \label{heat equation}
\end{equation}
admits an infinite dimensional group 
of Lie point symmetries generated by ,\cite{O-1993},
\numparts
\begin{equation}\label{heat finite algebra}
\eqalign{
\V_1  =  \partial_x,\qquad
\V_2  =  \partial_t,\qquad 
\V_3  =  u\partial_u,\qquad
\V_4  =  x\partial_x + 2t\partial_t, \\
\V_5  =  2t\partial_x - xu\partial_u,\qquad
\V_6  =  4tx\partial_x + 4t^2\partial_t - (x^2+2t)u\partial_u,}
\end{equation}
\begin{equation}
\V_\alpha = \alpha(x,t)\partial_u \qquad\text{where}\qquad
            \alpha_t=\alpha_{xx}.\label{heat infinite algebra}
\end{equation}
\endnumparts
To discretize \eref{heat equation}, we need 
two discrete indices, $(m,n)\in \mathbb{Z}^2$.

To perform our invariant discretisation procedure of the 
heat equation we consider only the six dimensional Lie algebra 
\eref{heat finite algebra}.  Before computing the 
invariants of \eref{heat finite algebra}, we point
out the important fact that
\begin{equation}
\Delta t_+=0\label{weak heat}
\end{equation}
is an invariant manifold of \eref{heat finite algebra}, 
which we will take to be one of 
our equations describing the lattice.  
Hence, the invariants are to
be computed on a grid with flat time layers.  
This restriction is important since 
we want to be able, at any time iteration,
to move everywhere in the spatial domain.  

The invariants are computed on a scheme involving
the points $(x,t,u)$, $(x_\pm,t,u_\pm)$ and $(\hat{x},\hat{t},\hat{u})$.
This is the minimum of point necessary to approximate the derivatives
present in the heat equation and generate 
an explicit scheme.  The set of elementary invariants is
obtained by solving the system of linear partial dif\mbox{}ferential 
equations
\begin{equation}
\pr\;\V_k[I(x,t,u,x_+,u_+,x_-,u_-,\hat{x},\hat{t},\hat{u})]=0,
\qquad k=1,\ldots,6,
\label{system invariant}
\end{equation}
where the quantity $I$ is supposed to depend on the discrete points
involved in the scheme and the
$\V_k$ are given in \eref{heat finite algebra}.
The solution of \eref{system invariant}, by the method of
characteristics,
gives the set of elementary invariants, \cite{V-2005},  
\begin{equation}\label{heat invariants}
\eqalign{
I_1=\frac{\Delta x_+}{\Delta x_-},\qquad
I_2=\frac{\Delta x_+^2}{\hat{\Delta t}}\left( \frac{u}{\hat{u}}\right)^2
\exp\left[\frac{-\hat{\Delta x}^2}{2\hat{\Delta t}}\right],\\
I_3=\frac{\Delta x_+^2}{4\hat{\Delta t}}-\frac{\Delta x_+^2}{\Delta x_+ +\Delta x_-}
\left\{\frac{1}{\Delta x_+}\ln\left(\frac{u_+}{u}\right)+
\frac{1}{\Delta x_-}\ln\left(\frac{u_-}{u}\right)\right\},\\
I_4=\frac{\Delta x_+\hat{\Delta x}}{\hat{\Delta t}}+\frac{2\Delta x_+}{\Delta x_+ +\Delta x_-}
\left\{\frac{\Delta x_-}{\Delta x_+}\ln\left(\frac{u_+}{u}\right)-
\frac{\Delta x_+}{\Delta x_-}\ln\left(\frac{u_-}{u}\right)\right\},\\} 
\end{equation}
The number of elementary invariants is coherent with the formula
\eref{number of invariants}.  Indeed, since the scheme
involves the points $(x,t,u)$, $(x_\pm,t,u_\pm)$ and 
$(\hat{x},\hat{t},\hat{u})$, the manifold $M$ introduce
in equation \eref{number of invariants} is given
by $M\sim \{ x,t,u,x_+,u_+,x_-,u_-,\hat{x},\hat{t},\hat{u}\}$
so dim $M$=10.  Furthermore, since the prolongation of
the 6 vector fields given in \eref{heat finite algebra}
are independent we get from \eref{number of invariants},
$\mu=10-6=4$.

From the set \eref{heat invariants} and the weakly 
invariant equation \eref{weak heat} we define the explicit
invariant scheme
\begin{equation}
I_2=4 I_3, \qquad \Delta t_+=0,\qquad I_4=0,
\end{equation}
which gives in terms of the original variables
\numparts
\begin{eqnarray}
\fl\left(\frac{u}{\hat{u}}\right)^2\exp\left[-
\frac{\hat{\Delta x}^2}{2\hat{\Delta t}}\right]
=1-\frac{4\hat{\Delta t}}{\Delta x_++\Delta x_-}\left\{\frac{1}{\Delta x_+}
\ln\left(\frac{u_+}{u}\right)+
\frac{1}{\Delta x_-}\ln\left(\frac{u_-}{u}\right)\right\},
\label{explicite discrete heat u}\\
\fl\Delta t_+=0,\label{explicite discrete heat t}\\
\fl\hat{\Delta x}= \frac {2\hat{\Delta t}}{\Delta x_++\Delta x_-}\left\{\frac{\Delta x_+}{\Delta x_-}
\ln\left(\frac{u_-}{u}\right)-\frac{\Delta x_-}{\Delta x_+}\
\ln\left(\frac{u_+}{u}\right)\right\}.
\label{explicite discrete heat x}
\end{eqnarray}
\endnumparts

Notice that the invariant scheme is not linear even though the 
PDE is.  This is due to the fact that we have neglected, 
in our derivation of
the discrete invariants \eref{heat invariants}, the infinite dimensional
symmetry generator \eref{heat infinite algebra}, 
which states that to any solution of the heat equation
we can add another solution.  

Let us prove that the system of equations
\eref{explicite discrete heat u}, \eref{explicite discrete heat t} and
\eref{explicite discrete heat x} is a valid
approximation of the heat equation \eref{heat equation} by computing
its continuous limit. 

We first start by taking the continuous limit in the discrete variable
$n$, meaning that $\Delta x_\pm$ and $\Delta t_+$ go to zero.  By doing
so, the equations \eref{explicite discrete heat u}, 
\eref{explicite discrete heat t} and
\eref{explicite discrete heat x} go to 
\numparts
\begin{eqnarray}
\left(\frac{u}{\hat{u}}\right)^2\exp\left[-
\frac{\hat{\Delta x}^2}{2\hat{\Delta t}}\right]
=1-2\hat{\Delta t}\left(\frac{u_{xx}}{u}-\left(\frac{u_x}{u}\right)^2\right),
\label{limit heat equation}\\
0=0,\\
\hat{\Delta x}=-2\hat{\Delta t}\frac{u_x}{u},\label{limit x}
\end{eqnarray}
\endnumparts
respectively.  The development of \eref{limit heat equation} in Taylor 
series in term of $\hat{\Delta x}$ and $\hat{\Delta t}$ gives, if
we keep only the terms that don't go to zero in the continuous limit,
\begin{equation}
-2\frac{\sigma}{\tau}\frac{u_x}{u}-2\frac{u_t}{u}-\frac{1}{2}
\frac{\sigma^2}{\tau^2}= -2\frac{u_{xx}}{u}+
2\left(\frac{u_x}{u}\right)^2,
\label{limit u}
\end{equation}
Using \eref{limit x}, we replace the occurances of $\hat{\Delta x}$ by 
$\hat{\Delta t}$ in \eref{limit u} and take the limit 
$\hat{\Delta t}$ and $\hat{\Delta x}$ to zero.  
Doing so, the equation \eref{limit u} goes to the heat equation
while \eref{limit x} goes to $0=0$.

It can be shown, \cite{V-2005}, that
\begin{eqnarray}
\fl u(x,t)=\frac{1}{\sqrt{4\pi t}}e^{-x^2/4t}, &\qquad  t=\tau \;m +t_0,
&\qquad x=(h\;n+x_0)(\tau\;m+t_0),\nonumber
\\
\fl u(x,t)=Ke^{-cx+c^2t}, &\qquad  t=\tau\; m+t_0, &\qquad x=h\;n+x_0+2c
(\tau\; m+t_0),
\label{heat solutions}
\end{eqnarray}
where $K$, $h$, $x_0$, $t_0$ $\tau$ and $c$ are constants, 
are non-trivial exact solutions of the invariant scheme composed of
the equations
\eref{explicite discrete heat u}, \eref{explicite discrete heat t} and
\eref{explicite discrete heat x}.    

Now let us perform the pure hodograph transformation 
$H:\mathbb{R}^3\to\mathbb{R}^3$, 
$(x,t,u)\mapsto (y=u,t,v=x)$.  Under such a transformation, 
the heat equation maps to the nonlinear equation
\begin{equation}
v_t=\frac{v_{yy}}{v^2_y}.\label{nonlinear heat}
\end{equation}
The symmetry generators of this equation are, \cite{O-1995}
\numparts
\begin{equation}
\eqalign{
\widetilde{\V}_1  =  \partial_v,\qquad 
\widetilde{\V}_2  =  \partial_t,\qquad 
\widetilde{\V}_3  =  y\partial_y,\qquad 
\widetilde{\V}_4  =  v\partial_v + 2t\partial_t, \\
\widetilde{\V}_5  =  2t\partial_v - vy\partial_y,\qquad 
\widetilde{\V}_6  =  4tv\partial_v + 4t^2\partial_t - 
(v^2+2t)y\partial_y,}
\label{heat finite algebra h}
\end{equation}
\begin{equation}
\widetilde{\V}_\alpha = \alpha(v,t)\partial_y 
\qquad\text{where}\qquad \alpha_t=\alpha_{vv}.
\end{equation}
\endnumparts
Clearly, since $t$ is unaf\mbox{}fected 
by the hodograph transformation,
the expression $\Delta t_+=0$ will remain an
invariant manifold of the new symmetry algebra. The discrete
invariants on such an invariant manifold are obtained directly from
\eref{heat invariants}, we just perform the hodograph
transformation on the set of invariants.  Hence we obtain, \cite{V-2005}
\begin{equation}
\fl\eqalign{
\widetilde{I}_1=\frac{v_+-v}{v-v_-},\qquad
\widetilde{I}_2=\frac{(v_+-v)^2}{\hat{\Delta t}}\left( 
\frac{y}{\hat{y}}\right)^2
\exp\left[\frac{-(\hat{v}-v)^2}{2\hat{\Delta t}}\right],\\
\widetilde{I}_3=\frac{(v_+-v)^2}{4\hat{\Delta t}}-\frac{(v_+-v)^2}{v_+ -v_-}
\left\{\frac{1}{v_+-v}\ln\left(\frac{y_+}{y}\right)+
\frac{1}{v-v_-}\ln\left(\frac{y_-}{y}\right)\right\},\\
\widetilde{I}_4=\frac{(v_+-v)(\hat{v}-v)}{\hat{\Delta t}}+
\frac{2(v_+-v)}{v_+ +v_-}
\left\{\frac{v-v_-}{v_+-v}\ln\left(\frac{y_+}{y}\right)-
\frac{v_+-v}{v-v_-}\ln\left(\frac{y_-}{y}\right)\right\},
}
\label{heat invariants h}
\end{equation}
Using the same invariant expressions as for the discrete
heat equation, i.e.
\begin{equation}
\widetilde{I}_2=4 \widetilde{I}_3,\qquad
\Delta t_+=0,\qquad \widetilde{I}_4=0,
\end{equation}
we get the explicit invariant scheme for \eref{nonlinear heat},
\numparts
\begin{eqnarray}
\fl\left(\frac{y}{\hat{y}}\right)^2\exp\left[-
\frac{(\hat{v}-v)^2}{2\hat{\Delta t}}\right]
=1-\frac{4\hat{\Delta t}}{v_+-v_-}\left\{\frac{1}{v_+-v}
\ln\left(\frac{y_+}{y}\right)+
\frac{1}{v-v_-}\ln\left(\frac{y_-}{y}\right)\right\},
\label{explicite discrete heat h v}\\
\fl\Delta t_+=0,\label{explicite discrete heat h t}\\
\fl\frac{\hat{v}-v}{\hat{\Delta t}}=\frac{2}{v_+-v_-}
\left\{\frac{v_+-v}{v-v_-}
\ln\left(\frac{y_-}{y}\right)-\frac{v-v_-}{v_+-v}
\ln\left(\frac{y_+}{y}\right)\right\}
\label{explicite discrete heat h y}
\end{eqnarray}
\endnumparts

The computation of the continuous limit of 
\eref{explicite discrete heat h v}, \eref{explicite discrete heat h t}
and \eref{explicite discrete heat h y}
is similar to the case of the heat equation.
Firstly, by letting the steps
generated by the discrete variable $n$ go to zero we get
respectively for equation 
\eref{explicite discrete heat h v}, \eref{explicite discrete heat h t}
and \eref{explicite discrete heat h y}
\numparts
\begin{eqnarray}
\left(\frac{y}{\hat{y}}\right)^2\exp\left[-
\frac{2\hat{\Delta t}}{y^2v_y^2}\right]=1+2\hat{\Delta t}
\left(\frac{v_{yy}}{yv_y^3}+\frac{1}{y^2v_y^2}\right)\\
0=0\\
\frac{\hat{v}-v}{\hat{\Delta t}}=-\frac{2}{yv_y}
\end{eqnarray}
\endnumparts
By developping the two nontrivial equations in Taylor series with
respect to the step generated by the discrete variable $m$ we get
\numparts
\begin{eqnarray}
\frac{\hat{\Delta y}}{\hat{\Delta t}}v_y+v_t+
\mathcal{O}(\frac{\hat{\Delta y^2}}{\hat{\Delta t}},\hat{\Delta t},
\hat{\Delta y})=-\frac{2}{yv_y}\label{limit tmp nonlinear heat 1}\\
\hat{\Delta y}=-\hat{\Delta t}\left(\frac{v_{yy}}{v_y^3}+\frac{2}{yv_y^2}
\right).\label{limit tmp nonlinear heat 2}
\end{eqnarray}
\endnumparts
By replacing the appearances of $\hat{\Delta y}$ in terms of $\hat{\Delta t}$
in \eref{limit tmp nonlinear heat 1} with the help of \eref{limit tmp nonlinear
heat 2} and then taking the limit $\hat{\Delta y}$ and $\hat{\Delta t}$
to zero we recover from \eref{limit tmp nonlinear heat 1} the PDE 
\eref{nonlinear heat} while \eref{limit tmp nonlinear heat 2} goes to 
$0=0$.

Finally, by a direct substitution 
we can verify that the exact solutions of the linear heat 
equation, \eref{heat solutions}, 
are again solutions of the invariant schemes
of the nonlinear partial dif\mbox{}ferential equation 
\eref{nonlinear heat}, after performing the
hodograph transformation on them.  Namely, we have that
\begin{eqnarray}
\fl v=(h\;n+v_0)(\tau\;m +t_0), &\qquad t=\tau\; m +t_0, &\qquad 
y=\frac{1}{\sqrt{4\pi t}}e^{-v^2/4t}, \nonumber
\\
\fl v=h\;n+v_0+2c(\tau\; m +t_0), &\qquad  t=\tau\; m+t_0, &\qquad y=Ke^{-cv+c^2t}, 
\end{eqnarray}
are exact solutions of 
the invariant system composed of equations 
\eref{explicite discrete heat h v}, \eref{explicite discrete heat h t}
and \eref{explicite discrete heat h y}, 
where $K$, $h$, $v_0$, $t_0$,
$\tau$ and $c$ are constants.

From this last example, we notice that even though 
a pure hodograph transformation can linearize a
dif\mbox{}ferential equation, it does not mean 
that the transformation on a discrete invariant
scheme will have the
same ef\mbox{}fect.  Indeed, the inverse 
hodograph transformation can 
be use to transform the nonlinear
equation \eref{nonlinear heat} to
\eref{heat equation}.  However, since the invariant scheme 
of the latter is not linear, the
discrete invariant scheme for $v_t=v_y^{-2}v_{yy}$ 
will not be linearized.   

\subsection{(1+1) dimensional wave equation with a source term}\label{wave}

In this section we obtain a symmetry-preserving
scheme of the (1+1) dimensional wave equation 
\begin{equation}
u_{tt}-u_{xx}=-4F(u),\label{wave equation}
\end{equation}
where $F(u)$ is an arbitrary function of $u$.  The symmetry
algebra of \eref{wave equation} is generated by
\begin{equation}
\V_1=t\p{x}+x\p{t},\qquad \V_2=\p{x},\qquad \V_3=\p{t}.
\label{symmetry algebra wave}
\end{equation}
instead of finding an invariant scheme in
the space $\{x,t,u\}$ we make the change of coordinates
to the characteristic variables
\begin{equation}
\psi:\mathbb{R}^2\to \mathbb{R}^2,\qquad
(x,t)\mapsto(y,z)=(x+t,x-t).
\label{caracteristic variables}
\end{equation}
Under the transformation \eref{caracteristic variables}, the 
PDE \eref{wave equation} is mapped to
\begin{equation}
u_{yz}=F(u).\label{transformed wave equation}
\end{equation}
A basis of the Lie symmetry algebra of \eref{transformed wave equation} is
\begin{equation}
\widetilde{\V}_1=y\p{y}-z\p{z},\qquad, \widetilde{\V}_2=\p{y},
\qquad \widetilde{\V}_3=\p{z},
\label{symmetry algebra caracteristic variables}
\end{equation}
where $\widetilde{\V}_2=1/2(d\psi(\V_2+\V_3))$ 
and $\widetilde{\V}_3=1/2(d\psi(\V_2-\V_3))$.

Before computing any invariants we have that
\begin{equation}
\hat{y}-y=0,\qquad z_+-z=0\label{weakly invariant equations wave}
\end{equation}
are two weakly invariant equations of the symmetry algebra 
\eref{symmetry algebra caracteristic variables} since they verify
the condition \eref{weakly invariance}.  By including these
two equations in our invariant scheme we can compute the
invariants on an orthogonal lattice, figure 2. 
\begin{figure}
\setlength{\unitlength}{0.7mm}
\begin{center}
\begin{picture}(120,60)
\put(0,10){\line(1,0){120}}
\put(0,30){\line(1,0){120}}
\put(0,50){\line(1,0){120}}
\put(20,0){\line(0,1){60}}
\put(60,0){\line(0,1){60}}
\put(100,0){\line(0,1){60}}
\put(20,30){\circle*{2.5}}
\put(60,30){\circle*{2.5}}
\put(100,30){\circle*{2.5}}
\put(60,10){\circle*{2.5}}
\put(60,30){\circle*{2.5}}
\put(60,50){\circle*{2.5}}
\put(100,50){\circle*{2.5}}
\put(62,52){$(y,\hat{z},\hat{u})$}
\put(62,32){$(y,z,u)$}
\put(62,12){$(y,\check{z},\check{u})$}
\put(102,32){$(y_+,z,u_+)$}
\put(22,32){$(y_-,z,u_-)$}
\put(102,52){$(y_+,\hat{z}_+,\hat{u}_+)$}
\end{picture}
\caption{Invariant lattice of equation \eref{transformed wave equation}.}
\end{center}
\label{lattice wave equation}
\end{figure}
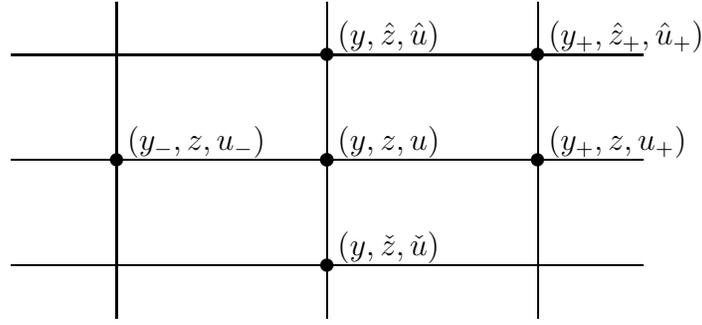 
A basis of elementary invariants in the space of discrete points
$\{(y,\hat{z},\hat{u}),$ $(y,z,u),$ $(y,\check{z},\check{u}),$
$(y_+,z,u_+),$ $(y_-,z,u_+),$ $(y_+,\hat{z}_+,\hat{u}_+)\}$
is
\begin{equation}
\eqalign{
\widetilde{I}_1=u,\quad \widetilde{I}_2=u_+,\quad 
\widetilde{I}_3=u_-,\quad \widetilde{I}_4=\check{u},
\widetilde{I}_5=\hat{u}, \quad \widetilde{I}_6=\hat{u}_+,\\
\widetilde{I}_7=\frac{\Delta y_+}{\Delta y_-},\quad
\widetilde{I}_8=\frac{\hat{\Delta y}_+}{\Delta y_+},\quad
\widetilde{I}_9=\frac{\hat{\Delta z}}{\check{\Delta z}},\quad
\widetilde{I}_{10}=\Delta y_+\hat{\Delta z},
}\label{invariants wave}
\end{equation}
where $\hat{\Delta y}_+=\hat{y}_+-\hat{y}$.

From this set of invariants we generate a symmetry-preserving
scheme by setting
\numparts
\begin{eqnarray}
\frac{\widetilde{I}_6-\widetilde{I}_5}{\widetilde{I}_8\widetilde{I}_{10}}-
\frac{\widetilde{I}_2-\widetilde{I}_1}{\widetilde{I}_{10}}=F(\widetilde{I}_1),
\label{solution wave}\\
\hat{y}-y=0,\qquad z_+-z=0.\label{lattice wave}
\end{eqnarray}
\endnumparts 
The equations for the lattice, \eref{lattice wave}, can be solved and
give
\begin{equation*}
y=\epsilon(m)\;n+y_0(m),\qquad z=\delta(n)\;m+z_0(n).
\end{equation*}
From \eref{invariants wave} 
we see that it is also possible to impose
$\Delta y_+=\Delta y_-$, by setting $\widetilde{I}_7=1$
and $\hat{\Delta z}=\check{\Delta z}$ by setting $\widetilde{I}=1$
without loosing any symmetries.  So if we
do so, the final solution for
the lattice is
\begin{equation}
y=\epsilon \; n+y_0,\qquad z=-\delta \; m-z_0,
\end{equation}
where $\epsilon$, $y_0$, $\delta$ and $z_0$ are constants.
On this rectangular lattice, the equation \eref{solution wave}
becomes in term of the discrete variables
\begin{equation}
\frac{\hat{u}_+-\hat{u}-u_++u}{\epsilon \delta }=F(u),
\end{equation}
which correspond to a standard discretisation of \eref{wave equation}.
The interesting thing to do now is to come back to the original
variables $\{x,t,u\}$ by applying the inverse transformation of
\eref{caracteristic variables}
\begin{equation}
\psi^{-1}(x,t)=(\frac{1}{2}(y+z),\frac{1}{2}(y-z)).
\label{inverse transformation wave}
\end{equation} 

First of all, the invariant equations \eref{weakly invariant equations wave} 
become
\begin{equation}
\hat{\Delta x}=-\hat{\Delta t},\qquad \Delta x_+=\Delta t_+.
\label{weakly invariant equations wave 2}
\end{equation}
The equations \eref{weakly invariant equations wave 2} are
weakly invariant equations of \eref{symmetry algebra wave} and
are chosen to be part of the invariant scheme of \eref{wave equation}.
The basis of invariants \eref{invariants wave} on the invariant
manifold \eref{weakly invariant equations wave 2}
becomes
\begin{equation}
\eqalign{
I_1=u,\quad I_2=u_+,\quad 
I_3=u_-,\quad I_4=\check{u},\quad
I_5=\hat{u}, \quad I_6=\hat{u}_+,\\
I_7=\frac{\Delta x_+}{\Delta x_-},\quad
I_8=\frac{\hat{\Delta x}_+}{\Delta x_+},\quad
I_9=\frac{\hat{\Delta t}}{\check{\Delta t}},\quad
I_{10}=-4\Delta x_+\hat{\Delta t},
}\label{invariants wave 2}
\end{equation}
under the transformation \eref{inverse transformation wave}.
By setting the same combination of invariants as in \eref{solution wave}
and \eref{lattice wave} and adding the equations $I_7=1$ and
$I_9=1$ we get an invariant scheme of \eref{wave equation}
\begin{equation}
\eqalign{
\frac{I_6-I_5}{I_8I_{10}}-
\frac{I_2-I_1}{I_{10}}=F(I_1),\\
\hat{\Delta x}=-\hat{\Delta t},\quad \Delta x_+=\Delta t_+,\quad
I_7=1,\quad I_9=1.}
\end{equation}
Which gives in term of the original variables
\numparts
\begin{eqnarray}
\frac{\hat{u}_+\hat{u}}{\hat{\Delta x}_+\hat{\Delta t}}-
\frac{u_+-u}{\Delta x_+ \hat{\Delta t}}=-4F(u),
\label{invariant discretisation wave}\\
\hat{\Delta x}=-\hat{\Delta t},\qquad \hat{\Delta t}=\check{\Delta t},\\
\Delta x_+=\Delta x_-,\qquad \Delta x_+=\Delta t_+,
\end{eqnarray}
\endnumparts
where $\hat{\Delta x}_+=\hat{x}_+-\hat{x}$.
The solution for the lattice is obtain from the known solution for the 
mesh in $y$ and $z$ and using the transformation \eref{inverse transformation
wave}
\begin{equation}
x=\frac{1}{2}(\epsilon \;n+y_0-\delta \; m-z_0),\qquad 
t=\frac{1}{2}(\epsilon \; n+y_0+\delta \; m+z_0).
\label{lattice wave xt}
\end{equation}
The solution for the mesh \eref{lattice wave xt} is drawn in figure
3. 

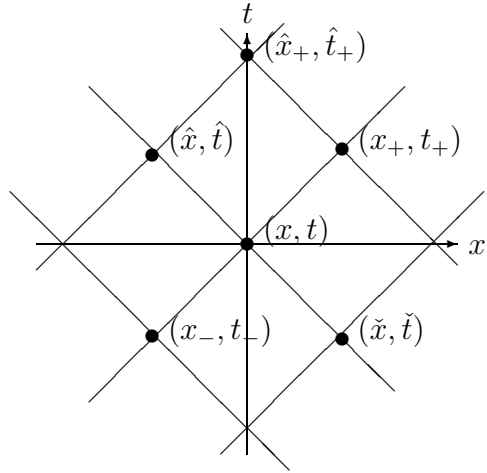
\begin{figure}
\setlength{\unitlength}{0.7mm}
\begin{center}
\begin{picture}(80,85)
\put(0,40){\vector(1,0){80}}
\put(40,0){\vector(0,1){80}}
\put(82,38){$x$}
\put(39,82){$t$}
\put(10,10){\line(1,1){60}}
\put(0,35){\line(1,1){47}}
\put(35,0){\line(1,1){50}}
\put(10,70){\line(1,-1){58}}
\put(35,81){\line(1,-1){45}}
\put(-5,50){\line(1,-1){53}}
\put(22,57){\circle*{2.5}}
\put(40,40){\circle*{2.5}}
\put(58,22){\circle*{2.5}}
\put(22,22.5){\circle*{2.5}}
\put(58,58){\circle*{2.5}}
\put(40,76){\circle*{2.5}}
\put(25,58){$(\hat{x},\hat{t})$}
\put(43,41){$(x,t)$}
\put(61,22){$(\check{x},\check{t})$}
\put(61,58){$(x_+,t_+)$}
\put(25,22){$(x_-,t_-)$}
\put(43,76){$(\hat{x}_+,\hat{t}_+)$}
\end{picture}\label{figure lattice wave xt}
\caption{Invariant lattice of equation \eref{wave equation}.}
\end{center}
\end{figure} 

We compute the continuous to show that the finite dif\mbox{}ference equation
\eref{invariant discretisation wave} is a valid approximation of
the partial dif\mbox{}ferential equation \eref{wave equation} 
on the lattice given by \eref{lattice wave xt}.  Taking into account
that the lattice implies
\begin{equation}
\Delta x_+=\Delta t_+\qquad \text{and}\qquad 
\hat{\Delta x}_+=\Delta x_+,
\end{equation}
we find by taking the Taylor series in the steps involving 
the variation of the discrete variable $n$ of \eref{invariant discretisation
wave} that
\begin{equation}
\eqalign{
\fl\left\{\hat{u}_x+\hat{u}_t+\frac{\Delta x_+}{2}(\hat{u}_{xx}
+\hat{u}_{tt}+2\hat{u}_{xt})+\mathcal{O}(\Delta x_+^2)\right.\\
\left.-\left(u_x+u_t+\frac{\Delta x_+}{2}(u_{xx}+u_{tt}+2u_{xt})+
\mathcal{O}(\Delta x_+^2)\right)\right\}\frac{1}{\hat{\Delta t}}=-4F(u),
}\label{limite tmp wave}
\end{equation}
where the hat over the derivatives of $u$ means that they are evaluated at
$(\hat{x},\hat{t})$.  By developping the hat derivatives around $(x,t)$
we get
\begin{equation}
u_{tt}-u_{xx}+\mathcal{O}(\Delta x_+,\hat{\Delta t})=-4f(u),
\end{equation}
which goes to the desired limit when $\Delta x_+$ and $\hat{\Delta t}$
go to zero.

The invariant discretisation \eref{invariant discretisation wave}
of \eref{wave equation} is quite dif\mbox{}ferent from the standard 
discretisation
\begin{equation*}
\eqalign{
\frac{u_+-2u+u_-}{\epsilon^2}-\frac{\hat{u}-2u+\check{u}}{\delta^2}
=-4F(u),\\
x=\epsilon \;n+ x_0,\qquad t=\delta \; m+t_0.}
\end{equation*}
The dif\mbox{}ference is due to the counterclockwise 
rotation of 45 degrees of the rectangular lattice
in the invariant case. 

\subsection{Polar coordinates transformation}\label{polar}

A more interesting transformation of coordinates
to be considered is the passage for cartesian 
to polar coordinates.  Unlike the other examples the
two spaces of variables are dif\mbox{}ferent since the
metric is not the same for each space.

We consider the elliptic partial dif\mbox{}ferential equation
\begin{equation}
u_{xx}+u_{yy}=F(u,x^2+y^2),
\label{laplace}
\end{equation}
where $F$ is an arbitrary function.  The only
symmetry of \eref{laplace} is the invariance
under rotation
\begin{equation}
\V=-y\p{x}+x\p{y}.\label{rotation generator}
\end{equation}
Nevertheless it is not obvious how to discretize \eref{laplace}
in cartesian coordinates and preserve the rotationnal symmetry.
The natural thing to do is to pass to polar coordinates
\begin{equation}
x=r \cos \theta,\qquad y=r \sin \theta.
\label{polar coordinates}
\end{equation}
In polar coordinates the equation \eref{laplace} becomes
\begin{equation}
u_{rr}+\frac{1}{r}u_r+\frac{1}{r^2}u_{\theta\theta}=F(u,r)
\label{laplace polar coordinates}
\end{equation}
and \eref{rotation generator} becomes the generator of translation
in $\theta$
\begin{equation}
\widetilde{\V}=\p{\theta}.
\end{equation}

Without going through the whole algorithm for finding an invariant scheme, it
is clear that the standard discretisation of \eref{laplace polar coordinates}
\begin{equation}
\eqalign{
\frac{u_+-2u+u_-}{\epsilon^2}+\frac{1}{r}\frac{u_+-u}{\epsilon}
+\frac{1}{r^2}\frac{\hat{u}-2u+\check{u}}{\delta^2}=F(u,r),\\
\hat{r}-r=0,\quad r_+-r=\epsilon,\quad \theta_+-\theta=0,\quad
\hat{\theta}-\theta=\delta,}
\label{invariant discretisation polar laplace equation}
\end{equation}
is invariant under translation in theta since it
only involves the dif\mbox{}ference between two values of
theta.  The solution for the lattice is given by
\begin{equation}
r=\epsilon \;n+r_0,\qquad \theta=\delta\;m+\theta_0.
\label{lattice polar laplace}
\end{equation}
where $\epsilon$,
$\delta$, $r_0$ and $\theta_0$ are constants.

Now, if we go back to cartesian coordinates 
\begin{equation*}
r=\sqrt{x^2+y^2},\qquad \theta=\arctan\left(\frac{y}{x}\right),
\end{equation*}
the
invariant scheme \eref{invariant discretisation polar laplace equation}
becomes
\begin{equation}
\fl\eqalign{
\frac{u_+-2u+u_-}{\epsilon^2}+\frac{1}{\sqrt{x^2+y^2}}\frac{u_+-u}{\epsilon}
+\frac{1}{x^2+y^2}\frac{\hat{u}-2u+\check{u}}{\delta^2}=
F(u,\sqrt{x^2+y^2}),\\
\sqrt{\hat{x}^2+\hat{y}^2}-\sqrt{x^2+y^2}=0,\quad 
\sqrt{x_+^2+y_+^2}-\sqrt{x^2+y^2}=\epsilon,\\
\arctan\bigg(\frac{y_+}{x_+}\bigg)-\arctan\bigg(\frac{y}{x}\bigg)=0,
\quad \arctan\bigg(\frac{\hat{y}}{\hat{x}}\bigg)
-\arctan\bigg(\frac{y}{x}\bigg)=\delta,}
\label{invariant discretisation laplace equation}
\end{equation}
and is invariant under \eref{rotation generator}.
From \eref{polar coordinates} and \eref{lattice polar laplace}
we have that the lattice in the $x$, $y$ variables is given by
\begin{equation}
\fl x=(\epsilon\;n+r_0)\cos(\delta \;m+\theta_0),\qquad
y=(\epsilon\;n+r_0)\sin(\delta\;m+\theta_0).
\label{invariant cartesian lattice laplace}
\end{equation}
In the Fig. 4 
we have illustrated
the lattice.  By carefully taking continuous limit of 
\eref{invariant discretisation laplace equation} on the
lattice \eref{invariant cartesian lattice laplace} we obtain,
as expected, the PDE \eref{laplace}.  

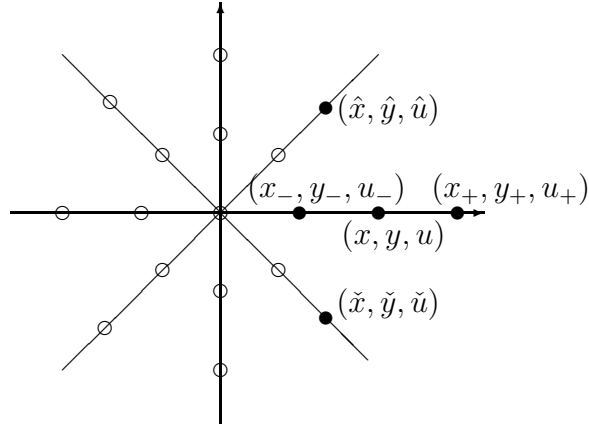
\begin{figure}
\setlength{\unitlength}{0.7mm}
\begin{center}
\begin{picture}(90,80)
\put(0,40){\vector(1,0){90}}
\put(40,0){\vector(0,1){80}}
\put(10,10){\line(1,1){60}}
\put(10,70){\line(1,-1){58}}
\put(40,40){\circle{2.5}}
\put(55,40){\circle*{2.5}}
\put(70,40){\circle*{2.5}}
\put(85,40){\circle*{2.5}}
\put(60,60){\circle*{2.5}}
\put(60,20){\circle*{2.5}}
\put(40,55){\circle{2.5}}
\put(40,70){\circle{2.5}}
\put(40,25){\circle{2.5}}
\put(40,10){\circle{2.5}}
\put(25,40){\circle{2.5}}
\put(10,40){\circle{2.5}}
\put(51,51){\circle{2.5}}
\put(29,29){\circle{2.5}}
\put(18,18){\circle{2.5}}
\put(51,29){\circle{2.5}}
\put(29,51){\circle{2.5}}
\put(19,61){\circle{2.5}}
\put(63,34){$(x,y,u)$}
\put(45,43){$(x_-,y_-,u_-)$}
\put(80,43){$(x_+,y_+,u_+)$}
\put(62,58){$(\hat{x},\hat{y},\hat{u})$}
\put(62,21){$(\check{x},\check{y},\check{u})$}
\end{picture}
\label{invariant lattice laplace}
\caption{Invariant lattice of equation \eref{laplace},\\
\begin{tabular}{cl}
\fullcircle & discrete points involved in 
\eref{invariant discretisation laplace equation}, \\
\opencircle & other discrete points.
\end{tabular}
}
\end{center}
\end{figure}

\section*{Conclusion}%

It is a well known fact that if the symmetry groups of two systems of
dif\mbox{}ferential equations are related by a point transformation
then these two systems are mapped into each other
by the same point transformation.  In this work, we have shown that
same result is true for invariant schemes.  That is, given
a system of dif\mbox{}ferential equations and its invariant scheme, any
point transformation will map the system of dif\mbox{}ferential equations
to a new system and the invariant scheme to a new invariant scheme
of the new system.  This result has been used to derive new invariant
schemes of dif\mbox{}ferential equations from known ones, 
section \ref{hodograph}.  We have obtained exact
solutions of these new schemes by applying the point transformation to
known solutions.  In section \ref{wave} and \ref{polar}, we have seen
that a change of coordinates can be used to facilitate the computation
of an invariant scheme of a dif\mbox{}ferential equation. 

The validity of the results obtained in this article rely heavily on
the point aspect of the transformation $\psi$.  
The result does not apply
to nonlocal transformations since such transformations will inevitably
break the local character of the symmetry transformations.
  
\ack
The author would like to thank NSERC for their financial support and
Pavel Winternitz for his judicious comments while this work was 
realized.   
  
\section*{References}%


\end{document}